\documentclass[twocolumn,showpacs,preprintnumbers,amsmath,amssymb,floatfix]{revtex4}

\usepackage{graphicx}
\usepackage{dcolumn}
\usepackage{bm}

\begin{document}


\title{Superfluid toroidal currents in atomic  condensates}

\author{Eileen Nugent $^{\dag}$}
\email{e.nugent@am.qub.ac.uk}
\author{Dermot McPeake $^{\ddag \dag}$}
\author{J F McCann  $^{\dag}$}

\address{
 $^{\dag}$ Department  of Applied Mathematics and Theoretical Physics,\\
Queen's University Belfast, Belfast BT7 1NN, Northern Ireland \\
$^{\ddag}$ NMRC, University College Cork,
Lee Maltings,
Prospect Row,
Cork,
Ireland.
}

\date{April 2003}

\begin{abstract}
The dynamics of toroidal condensates in the presence of condensate flow
and dipole perturbation have been investigated. The 
Bogoliubov spectrum of 
condensate is calculated for an oblate torus using a discrete-variable 
representation and a spectral method to high accuracy. The transition from 
spheroidal to toroidal geometry of the trap displaces 
the energy levels into narrow bands. 
The lowest-order acoustic modes are quantized with 
the dispersion  relation $\omega \sim |m| \omega_s $ with $m=0,\pm 1,\pm 2, \dots$.
A condensate with  toroidal current  $\kappa$ splits the 
$|m|$ co-rotating and counter-rotating pair by the amount: 
$\Delta E \approx 2 |m|\hbar^2 \kappa \langle r^{-2} \rangle$.
Radial dipole excitations are the lowest energy dissipation modes. 
For highly occupied condensates the nonlinearity creates an asymmetric 
mix of dipole circulation and nonlinear shifts in the spectrum of 
excitations so that the center of mass circulates around the 
axis of symmetry of the trap. We outline an experimental method 
to study these excitations.
\end{abstract}

\pacs{03.75.Kk,03.75.Lm}
\maketitle

\section{Introduction}
\label{sec1}

The elementary excitations of trapped Bose Einstein condensates 
have  been extensively studied in recent years. These collective modes 
are coherent macroscopic matter waves that can be used in many applications 
in cold atom Physics. Since the trap geometry defines the mode spectrum 
and amplitudes, recent studies have considered spheroidal condensates 
\cite{penc01,hutc98}  and in novel topologies such as toroidal 
traps \cite{rokh97,sala99,bena99}. 
A toroidal trap is of particular interest as it can be employed 
as a storage ring for coherent atom waves \cite{arno02} 
or ultracold molecules \cite{crom01}
enabling investigations of persistent currents,  Josephson effects, phase 
fluctuations
 and  high-precision Sagnac or gravitational interferometry. 
More adventurous possibilities for toroidal condensates include the construction 
of a mode-locked atom laser \cite{drum01} and the creation of sonic 
black holes in tight ring-shaped condensates \cite{gara01}. 
A narrow ring of condensate, effectively 
reducing the dimensionality to 1D, 
could be applied to dark soliton propagation \cite{bran01}
or low-dimensional quantum degeneracy 
including the Tonks gas regime \cite{petr00,gira01}. 
Toroidal traps have been around for some time, being employed in early 
experiments with Sodium vapors \cite{davi95}. In this case a blue-detuned 
laser was used  as a measure to 
counteract Majorana spin flips; a loss 
mechanism which can be problematic in the cooling required 
for condensation. In more recent experiments the toroidal topology has been used
in the study of vortex nucleation and superfluidity \cite{matt99}.  

Recent theoretical work concerning toroidal condensates has concentrated on 
excitations in traps which have some time dependence in their 
topology \cite{busc01} and on vortex-vortex dynamics which have
been shown to be strongly distorted by such a geometry \cite{schu02}.
The stability of multiply-quantized toroidal currents has been studied 
by \cite{busc01,bena99}. The spectrum of single-particle excitations 
for cigar-shaped toroidal traps with circulation has also been 
considered \cite{sala99}.
 Although a wide variety of toroidal trap parameters is possible, 
one of the advantages of such a system, the most easily accessible experiments 
appear to be based on oblate shaped traps. In this paper we study  
results for the 
spectrum of collective excitations of oblate toroidal condensates 
within the Bogoliubov approximation, and 
explores the dynamical stability of ring currents around the torus.
The main features we note are generic to this design of trap and would 
apply to similar geometries.
Perturbations of superfluid ring currents by an off-set of the 
trapping potential are studied.  For example, in a toroidal trap the central repulsive potential 
acts a pinning site for the vortex and thus the stability of the 
flow can be studied. 
 The excitation spectrum, mode
 densities, flow rate and center-of-mass motion  for 
this system are obtained by employing both a
 time-dependent  and time-independent  method \cite{mcpe02,nils03}. 
 A simple, but accurate,  formula is presented for the lowest angular acoustic 
 modes of excitation, and the splitting energy of these modes when a background
 current is present.

\section{The model}
\label{sec2}

\subsection{The weakly-interacting Bose gas}

For a cold dilute weakly-interacting gas, the ground state 
(condensate mode) dominates the collective
dynamics of the system. In experimental realizations one can
achieve temperatures such that $T \ll T_c$ (typically $0.1$ to
$1\mu{\rm K}$) and densities such that the gas is weakly 
interacting and highly dilute.  Under such conditions, the
condensate of $N_0 \gg 1$ atoms is well described by a mean field, or  
wavefunction, governed by the Gross-Pitaevskii equation, and the quasiparticle 
excitations are acoustic waves within this field. If the perturbations of the condensate 
are small, then it is appropriate and convenient to use the linear response
approximation, which is equivalent to the Bogoliubov  approximation 
for single-particle excitations in highly-condensed quantized Bose gases at zero
temperature.

Consider a dilute system of $N_0$ atoms, each of mass $m_a$, trapped by an 
external potential $V_{\rm ext}(\bm{x},t)$ and interacting weakly
through the two-body potential $V(\bm{x},\bm{x}')$.
At low temperatures and densities, the atom-atom interaction
can be represented perturbatively by the $s$-wave 
pseudopotential:  $V(\bm {x},\bm{x}')=  
(4\pi\hbar^{2}a_s/m_a)\delta ^{(3)}(\bm{x}
 - \bm{x}')$, and 
 $a_s$ is  the $s$-wave scattering length. The dynamics 
follow from the Hartree variational principle:
\begin{equation}
\delta \int_{t_1}^{t_2}  dt \int d^3 \bm{x} \ \ \psi^*[ H_0 +
\textstyle{1 \over 2} g \psi^* \psi-i\hbar\partial_t ]\psi =0
\label{vp}
\end{equation}
where $g=(4\pi \hbar^2 / m_a) N_0 a_s$, 
$H_0 = -(\hbar^2/2m_a)\nabla^2+V_{\rm ext}-\mu$, and 
the chemical potential $\mu$ plays the role of a Lagrange multiplier.
Supposing that the temperatures are sufficiently low that 
the condensate can be represented by a Bogoliubov mean field $\phi$. 
The single-particle excitations 
can be described by the linear response ansatz:
\begin{eqnarray}
\psi(\bm{x},t) & = &    a_0(t) \phi (\bm{ x}) + \\ \nonumber
   & & \sum_{j\neq 0} \left[ a_j(t)  {u}_j(\bm{x}) e^{-i\omega_j t}
+ a^*_j(t) {v}^*_j (\bm{x}) e^{+i\omega_jt}\right] 
\label{ansatz}
\end{eqnarray}
where $\phi$ represents the highly-occupied condensate; that is,
$ | a_0| \approx \sqrt{N} \gg |a_j|$, $j \neq 0$.
From the variation $\delta \phi^*$, and 
linear expansion  in the small parameters $a_j,a_j^*$ taken as 
constant, the stationary Gross-Pitaevskii equation and 
Bogoliubov equations follow:
\begin{equation}
H_0 \phi +g |\phi|^2\phi =0
\label{gpe}
\end{equation}
with $ (\phi, \phi) =1$  and $ \int d^3{\bm
x}\; f({\bm x})^* g({\bm x}) \equiv (f,g)$. 
The Bogoliubov modes are solutions of the
coupled linear equations:
\begin{eqnarray}
(H_0+2g |\phi|^2) u_j +g \phi^2\  v_j & = & +\hbar\omega_j u_j 
\label{bdg1} \\
(H_0+2g |\phi|^2)v_j +g \phi^{*2} u_j & = & -\hbar\omega_j v_j 
\label{bdg2}
\end{eqnarray}
Time-reversal symmetry of equations (\ref{bdg1},\ref{bdg2}) is reflected in
the fact that every set of solutions $\{\omega_j,u_{j},v_{j}\}$
has a corresponding set $\{-\omega_j,v^{\ast}_{j},u^{\ast}_{j}\}$ 
and the  normalization convention is:
$ (u_i,u_j)-(v_i,v_j) =\delta_{ij}$.

\begin{figure}[htbp]
\centering
\includegraphics[width=8cm]{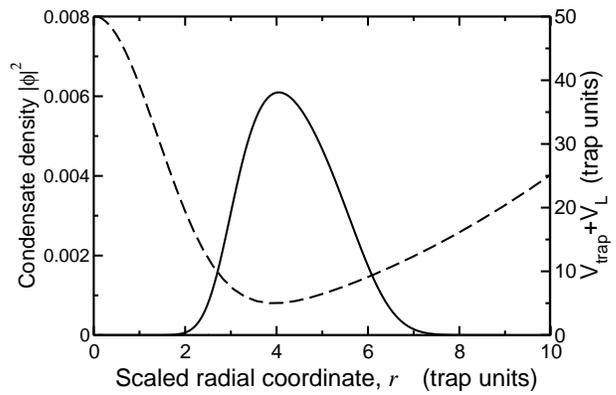} 
\caption{Torus potential (dashed line) and condensate density $\phi|^2$ 
(full line) as a function of radius 
in the  the plane $z=0$. 
The radial potential corresponds 
to  $ h_0=50 \hbar \omega_r$, and $\sigma = 2 \sqrt{\hbar/2 m_a\omega_r}$. 
The condensate circulation is $\kappa=0$ and the interaction 
strength is $C=1000$. The scaled condensate probability density $|\phi|^2$ is 
given in
units $({\hbar/2 m_a\omega_r})^{3/2}$
}
\label{figure1}
\end{figure}

\subsection{Toroidal condensates}

Conventional atom traps 
provide a confinement of the condensate in the radial and 
axial directions. We write this as a spheroidal  potential 
of the type: $V = {1 \over 2} m_a \omega_r^2(r^2+\lambda^2 z^2)$, 
where $r$ is the radial coordinate, $\omega_r$ the radial frequency, 
 and  $z$ the axial 
coordinate. Here $\lambda$ is the aspect ratio of the 
harmonic potential, so that  $\lambda >1$ flattens the 
condensate into an oblate pancake shape. 
For convenience we use scaled dimensionless units for length, time
and energy namely: $(\hbar/2m_a\omega_r)^{1 \over
2},\omega_r^{-1}$ and $ \hbar \omega_r$, respectively. 
Typically,  $\omega_r \sim 2\pi \times 200$Hz, so that
the natural timescale for oscillation is of the order of milliseconds.
If this potential is supplemented by a repulsive core then the 
toroidal shape can be realized. The conventional method is to use 
 blue-detuned laser light to create a dipole force. In the 
 experiment of Davis et al.  \cite{davi95} the AC Stark shift 
from the green light of an argon-ion laser ($\lambda=514$nm), detuned from the 
$D$-line  ($\lambda=589$nm) of the trapped sodium atoms created the dipole force. 
If the laser beam is aligned along the $z$-axis of the trap at the 
diffraction limit focus then the potential if a function of $r$ only:
$U(r)= \hbar I(r) (8 \tau^2\ \Delta . I_{sat})^{-1}$,
where  $I_{sat}$ is the saturation intensity of the level, $\Delta$ is the 
detuning 
and $\tau$ the lifetime of the upper atomic level \cite{adam97}. Across the 
focus the
intensity variation is Gaussian: $I(r) =I_0 \exp(-r^2/\sigma^2)$, where the 
on-axis intensity is related to the power $P$ and waist of the beam $w_0$ by 
$I_0=2P/\pi w_0^2$. So the width and height of the central barrier 
can be controlled by the these parameters. Consider a typical case for 
a trap containing sodium atoms; an 
argon-ion laser beam of 3.5 W focused to a waist  $w_0=30 \mu$m, 
would give a frequency shift of 7 MHz  at $r=0$. Compared with 
trap frequencies of order $\omega_r/2\pi \sim 100-400$Hz, this would be
sufficient to create a hole along  the trap axis and form a toroidal 
condensate.  The combination of the fields gives the external potential:
$V_{\rm ext} ({\bm x},0) = {1 \over 2} m_a \omega_r^2
(r^2+\lambda^2 z^2) + h_0 \exp(-r^2/\sigma^2)$, 
where the repulsive central barrier 
is defined by height and width parameters $h_0$ and $\sigma$, 
respectively. The radial minimum of the well is displaced to 
$r_0 = \sigma \sqrt{ \ln(4h_0/\sigma^2)}$. 
For convenience we use scaled dimensionless units for length, time
and energy namely: $(\hbar/2m_a\omega_r)^{1 \over
2},\omega_r^{-1}$ and $ \hbar \omega_r$, respectively 
Then it is convenient to define the interaction  strength by the 
dimensionless parameter $C \equiv  8\pi N_0 a_s 
 (\hbar/2m_a\omega_r)^{-{1 \over
2}}$, so that as millions of atoms occupy  the condensate 
$N_0 \rightarrow \infty, C \rightarrow \infty$, 
while the ideal gas corresponds to $C \rightarrow 0$. For the pancake geometry 
$\lambda \gg 1$ the 
central barrier height $h_0$ controls the transition from spheroid to 
torus. The radius $r_0$ of the potential well minimum provides 
a guide to the size of the torus.  Collective excitations for 
a quartic toroidal potential of the form: 
$ V(r,z)_{\rm ext} = {1 \over 2} m_a \omega_r^2
[(r-r_0)^2+\lambda^2 z^2]$ were considered by 
Salasnich et al \cite{sala99} and the variation in 
chemical potential and frequency with respect to  particle number 
was studied. The geometry chosen \cite{sala99} was a prolate 
shape such that $ \lambda=1/\sqrt{8}$ in contrast to the oblate case 
considered here. 
An example of a toroidal condensate 
density in  static equilibrium  is shown in figure  \ref{figure1}. 
The trap parameters are   $ h_0=50 \hbar \omega_r$, and 
$\sigma = 2 \sqrt{\hbar/2 m_a\omega_r}$ so that $r_0 \approx 4 $, and 
the interaction strength is $C=1000$. For strong interaction, high $C$,   
the radius of gyration ($r_K$) about the symmetry axis is a better 
estimate of condensate radius; where $ r_K^2  \equiv   
\int |\phi|^2 r^2 d^3 \bm{x}/\int |\phi|^2 d^3 \bm{x} $. 
For the pancake shape ($h_0 = 0$)
 $ r_K  =  3.2 $, but as the barrier is raised 
to $h_0 = 10 $ then there is some depletion of condensate at the center, so 
that it expands to $  r_K =  3.8$. Finally for
a high central barrier, $h_0 = 50 $,  the condensate is excluded 
from the trap axis and a narrow ring is formed with
$ r_K = 4.6 $ as shown in figure \ref{figure1}. Here the maximum 
condensate density is located close to the potential energy minimum 
 $r_0 \approx 4$ in accord with hydrostatic equilibrium \cite{dalf99}.

\begin{figure}[htbp]
\begin{center}
\includegraphics[width=7cm,angle=0,clip=on]{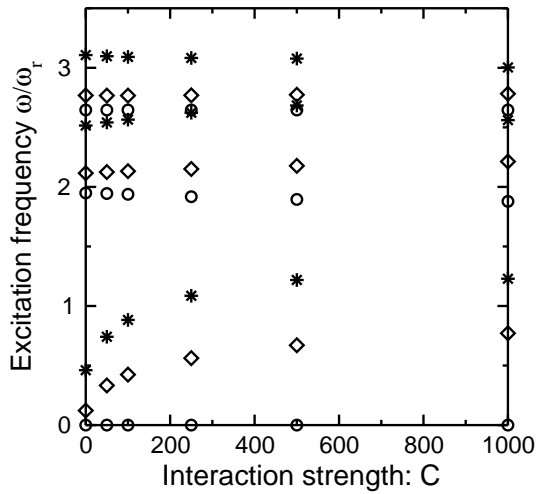} 
\caption{
Excitation spectrum for the axisymmetric modes
of semi-toroidal condensate without circulation ($\kappa=0$)  
as a function of interaction strength $C$. 
Trap parameters are
$\lambda = \sqrt{7}$, $\sigma = 2$.
The symmetries of the modes are: $m = 0\ (\circ)\ $,$|m|=1\ (\diamond)\ $,
$ |m|=2\ (\ast)\ $.   
}
\label{figure2}
\end{center}
\end{figure}

\begin{figure}[htbp]
\begin{center}
\includegraphics[width=7cm,angle=0]{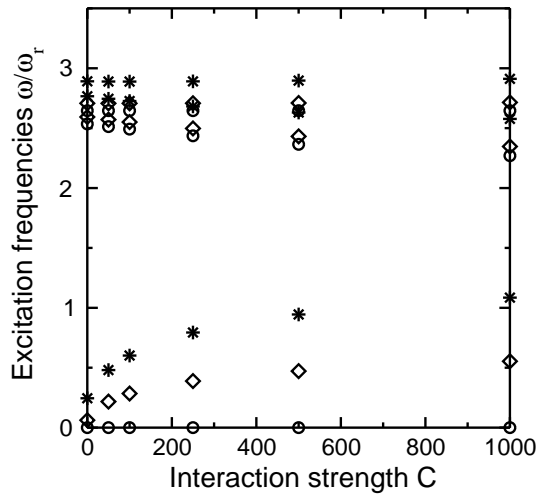} 
\caption{
Excitation spectrum for the axisymmetric  modes
of toroidal condensate without circulation ($\kappa=0$)
as a function of interaction strength $C$. 
Trap parameters are $\lambda = \sqrt{7}$, 
$\sigma = 2$ and $h_0=50$. 
The mode symmetries are: $m = 0\ (\circ)$,$|m|=1 \ (\diamond)$,
$ |m|=2\ (\ast)$ }
\label{figure3}
\end{center}
\end{figure}

\subsection{Bogoliubov spectrum}
\label{sec3}

 In this 
paper we are primarily interested in the low-lying axisymmetric 
excitations arising from weak perturbation of the
condensate, so that detailed results will be presented for the first 
few monopole, dipole  and 
quadrupole excitations only.
For finite $C$, the spectrum of excitations 
must be determined by numerical solution of 
equations (\ref{gpe},\ref{bdg1}) and 
(\ref{bdg2}). Separating variables gives:
\begin{equation}
\phi(r,z,\varphi) = \tilde{\phi}_{\kappa}(r,z)e^{i\kappa\varphi}
\end{equation}
so that the condensate, with circulation $\kappa$ and 
real amplitude $\tilde{\phi}$,  is the solution of the
equation:
\begin{eqnarray}
{\cal{L}}_{(\kappa)}  \tilde{\phi}_{\kappa}-g |\tilde{\phi}_{\kappa}|^{2}
\tilde{\phi}_{\kappa}= 0
\label{condeq}
\end{eqnarray}
where 
\begin{eqnarray}
{\cal{L}}_{(s)} \equiv 
- {\displaystyle {\hbar^2 \over 2m_a} \left( {\partial^2 \over \partial r^2}
+ \frac{1}{r} \frac{\partial}{\partial r}
+ \frac{\partial^{2}}{\partial z^{2}} 
- \frac{s^{2}}{r^{2}} \right)+  } \\ \nonumber
  {\textstyle {1 \over 2}}m_a\omega_r^2(r^2+\lambda^2 z^2)+h_0 
\exp(-r^2/\sigma^2) + 2g |\tilde{\phi}_{\kappa}|^{2}-\mu
\end{eqnarray}
The cylindrical symmetry of the condensate means that small amplitude excitations 
can be described by radial ($n_r$), axial ($n_z$) and rotational ($m$) quantum numbers, 
each with an associated  parity.
Here $m=n_{\theta}-\kappa=0,\pm 1,\pm 2, \dots$   denotes 
angular momentum with respect to the condensate, while $n_r,n_z=0,1,2,3,\dots $. 
The quasiparticle amplitudes:
\begin{eqnarray}
   u_{n_r,m,n_{z}}(r,z,\varphi) \equiv
\tilde{u}_{n_r,n_z}(r,z) e^{i(m + \kappa)\varphi} \\  
v_{n_r,m,n_z}(r,z,\varphi) \equiv
\tilde{v}_{n_r,n_z}(r,z) e^{i(m - \kappa)\varphi}
\end{eqnarray}
with corresponding angular frequency,
$\omega_{n_r,m,n_{z}}$,
are solutions of the eigenvalue problem
\begin{eqnarray}
{\cal{L}}_{(m+\kappa)}\tilde{u}(r,z) +
g\ \tilde{\phi}_{\kappa}^{2} \ \tilde{v}(r,z)
&=& +\hbar \omega \tilde{u}(r,z)  \label{dimbdg1}
\label{bogo1} \\
{\cal{L}}_{(m-\kappa)}\tilde{v}(r,z) +
g\ \tilde{\phi}_{\kappa}^{2}\ \tilde{u}(r,z)
&=& -\hbar \omega \tilde{v}(r,z) \label{dimbdg2}
\label{bogo2}
\end{eqnarray}

The equations are discretised  on a 
2D-grid using Lagrange functions  \cite{mcpe02}; 
 the radial coordinate is defined at $M$  grid points
($r_1,r_2,\dots,r_{M}$)
and the axial coordinate at $N$ points
($z_1,z_2,\dots,z_{N}$). Therefore:
\begin{equation}
\tilde{\phi}_{\kappa}(r,z) = \sum_{k=1}^{M}\sum_{l=1}^{N}
\tilde{\phi}_{\kappa}^{kl} (r_{k},z_{l}) \lambda_{k}^{-1/2} \mu_{l}^{-1/2}\
f_{k}(r)g_{l}(z)
\label{3ddvr}
\end{equation}
where $f,g$ are Lagrangian interpolating functions such that
\begin{eqnarray}
\int_{0}^{\infty} f_{i}^{\ast}(r)f_{k}(r)\ 2\pi r \ dr \approx
\lambda_{i}\delta_{ik}  \\
\int_{-\infty}^{\infty}g_{j}^{\ast}(z)g_{l}(z)\ dz \approx \mu_{j}\delta_{jl}
\end{eqnarray}
The functions for the $r$-coordinate are 
chosen to be generalized Laguerre polynomials \cite{mcpe02}, 
scaled to encompass the entire condensate, with typically
$M =50$ mesh points. On the other hand, Hermite polynomials are used
in the $z$-direction so that
\begin{equation}
g_{l}(z) = \sum_{l=0}^{N-1} \chi_{l}^{\ast} (z_{l})  \chi_{l}(z)
\end{equation}
where $\chi_{l}(z) = h_{N}^{-\frac{1}{2}} w(z)^{\frac{1}{2}} H_{l}(z)$.
 and $H_{l}(z)$ are the Hermite polynomials associated with weights
 $w(z)=e^{-z^2}$  and normalization factor
 $h_{N} = 2^{N}\pi^{1/2}N !$.
 A high degree of accuracy was found with only $N=30$ points.
The resultant eigenvalue problem  was
solved using the following approach. 
The condensate density was found using Newton's method for equation (\ref{condeq}). The
decoupled linear eigenvalue problem (\ref{dimbdg1},\ref{dimbdg2})
was solved by conversion to Hessenberg $QR$ form followed by 
inverse iteration \cite{golu96}.
Convergence was established by a combination of grid scaling
and number of mesh points, so that at least six-figure precision was 
assured for all frequencies (see table \ref{table1} and table \ref{table2}).

\begin{figure}[!tbp]
\begin{center}
\includegraphics[width= 8cm]{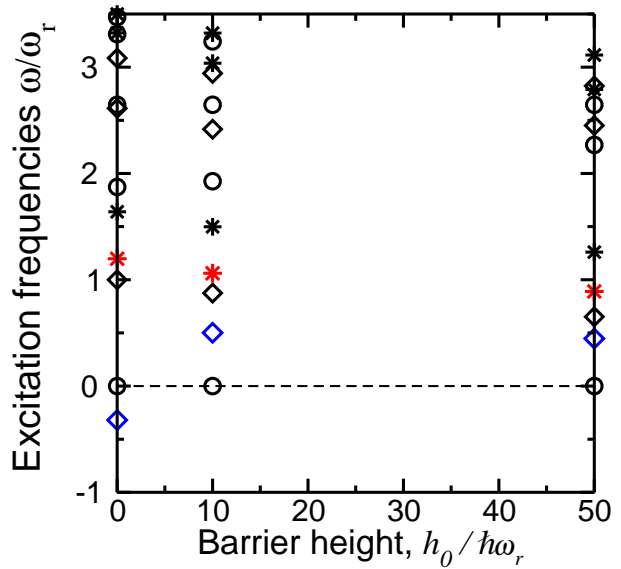}
\caption{\label{figure4}  
Excitation spectrum
 for modes of a condensate with circulation $\kappa = 1$.  
 The mode symmetries are $|m| = 0(\circ),1(\diamond),2(\ast)$. 
 Frequencies are plotted  as a 
function of barrier height $h_0$ 
in a trap characterized by $\lambda = \sqrt{7}$ 
and  $\sigma = 2 \sqrt{\hbar/2 m_a\omega_r}$, with interaction strength $C= 1000$.
See also table \ref{table2}} 
\end{center}
\end{figure}

\begin{table}
\caption{\label{table1}
Angular frequencies $\omega/\omega_r$ of the first-excited 
radial dipole modes. The table compares
the numerical accuracy of the  time-dependent 
linear response method  (TDLR) with $\epsilon =0.02$ 
with the time-independent methods (BdG) and agreement is better 
than 2\% in all cases. 
The condensate interaction strength is $C \equiv  8\pi N_0 a_s 
 (\hbar/2m_a\omega_r)^{-{1 \over
2}}=1000$ and the axial mode is $n_z=0$. 
Results are presented for $\kappa=0$ and $\kappa=1$.}
\begin{center}
\begin{ruledtabular}
\begin{tabular}{l|cccc}
 & \multicolumn{2}{c}{$h_{0}/\hbar\omega_r =10$}&\multicolumn{2}{c}
 {$h_{0}/\hbar\omega_r =50$} \\
mode & TDLR & BdG &  TDLR & BdG \\
\hline
$\kappa=0$  &  &  &  &   \\
$m=+1$  & 0.772 & 0.771  & 0.563 &  0.554 \\
\hline
$\kappa=0$  &  &  &  &   \\
$m=-1$ & 0.772 & 0.771  & 0.563 &  0.554 \\
 \hline
$\kappa=1$  &  &  &  &   \\
$m=+1$  & 0.875 & 0.874  & 0.657 & 0.652\\
\hline
$\kappa=1$  &  &  &  &   \\
$m=-1$ & 0.496 & 0.501  & 0.448 & 0.447 \\
\end{tabular}
\end{ruledtabular}
\end{center}
\end{table}

\begin{table}
\caption{\label{table2}
Excitation frequencies $\omega/\omega_r$ of the toroidal condensate 
with circulation $\kappa=1$ with variation in the central barrier 
height $h_0$.
The condensate interaction strength is $C \equiv  8\pi N_0 a_s 
 (\hbar/2m_a\omega_r)^{-{1 \over
2}}=1000$ }
\begin{center}
\begin{ruledtabular}
\begin{tabular}{l|ccc}
$m$   & \multicolumn{1}{c}{$h_{0}/\hbar\omega_r =0$} &
 \multicolumn{1}{c}{$h_{0}/\hbar\omega_r =10$}&\multicolumn{1}{c}
 {$h_{0}/\hbar\omega_r =50$} \\
&  2.646   &  2.646	&  2.646 \\
&  3.310   &  3.243	&  3.954\\
$m=0$ &  3.471   &  3.600	&  3.970 \\
&  4.691   &  4.480	&  4.771 \\
&  4.728   &  4.807	&  5.609 \\
&  4.936   &  5.027	&  5.882 \\
&	    &		&	  \\
& -0.320   &  0.501	&  0.447 \\
$|m|=1$ &  1.000   &  0.875	&  0.652 \\
&  2.613   &  2.417	&  2.451 \\
&  3.085   &  2.942	&  2.823 \\
&	    &		& 	  \\
&  1.197   &  1.061	&  0.890 \\
&  1.640   &  1.498	&  1.260 \\
$|m|=2$ &  3.499   &  3.037	&  2.790 \\
&  3.322   &  3.321	&  3.114 \\
&  4.925   &  4.530	&  4.396 \\
&  4.840   &  4.561	&  4.471 \\
\end{tabular}
\end{ruledtabular}
\end{center}
\end{table}

\subsection{Spectrum of torus excitations}
\label{sec3}

The Bogoliubov mode frequencies for a toroidal condensate without circulation 
($\kappa=0$) are shown in figures ~\ref{figure2} and ~\ref{figure3} for 
two different barrier heights. The variation in frequencies as the particle 
number ($C$) increases  is shown. In the absence of circulation for the low 
barrier $h_0=10$ (figure \ref{figure2}) the effect of increasing the interaction strength, 
 or equivalently increasing the number of atoms in the condensate, 
is to spread the spectral lines of the low-lying modes, while the highly-excited 
state frequencies are relatively insensitive as $C$ increases. 
In contrast, to the prolate toroid results  
\cite{sala99} the effect of increasing atom number, increasing $C$,leads to
an increase in the gap between the low-frequency excitations. 
The three $m=0$ modes 
shown are, in increasing frequency, the gapless mode $\omega=0$, the first breathing mode, 
and the first axial dipole mode $\omega=\sqrt{7}$. The lowest degenerate modes $m=\pm 1$ 
are the fundamental radial dipoles, at higher frequency $\omega \sim 2$ we observe an octupole mode, 
and finally the $xz$-quadrupole at $\omega \sim 2.8 $ is very close to the axial dipole.  
The increasing density of states for higher excitations 
 is familiar from studies of spherical condensates 
and is reproduced here. 
Finally the lowest pair of $m=2$ states, corresponding to 
quadrupole excitation in the radial coordinate  
are far below the next higher excitations of this symmetry. 

As the toroidal shape is more sharply defined ($h_0=50$) the 
high-frequency modes become more tightly grouped (figure \ref{figure2}) and the gap 
to the lowest-order excitations widens.  The long-wavelength low-energy modes are circulations
around the torus, as shown in figure  \ref{figure3}. The acoustic modes \cite{dalf99} 
appropriate in the limit $C \rightarrow \infty $, are sinusoidal 
oscillations in density $\delta\rho$ and follow the microscopic wave equation, in our units:
\begin{equation}
-\omega^2\delta\rho= 2 \nabla \cdot( [ \mu -V_{\rm ext} ] \nabla \delta\rho)
\label{acoustic}
\end{equation}
In the Thomas-Fermi limit, the condensate occupies the region $\Omega$ 
where $\mu -V_{\rm ext} >0$. The angular waves satisfy the periodic 
boundary conditions $\delta \rho (\varphi+2\pi m) = \delta \rho (\varphi)$. 
Taking the volume average of equation (\ref{acoustic}) over the radial and 
axial density gives: $\omega= |m| \omega_s$ where
\begin{equation}
\omega_s^2 \int_{\Omega} 2\pi r dr dz  =  2 C \int_{\Omega}
  r^{-2}\phi^2 2\pi r dr dz 
\end{equation}
So that:
\begin{equation}  
  \omega = |m| \sqrt{ 2C\langle \phi^2 r^{-2} \rangle }
\label{sound} 
\end{equation}
The results in figure \ref{figure2} reflect the change in 
dispersion relation  $\omega \sim m^2 $ to 
$\omega \sim m$ for the lower frequency modes as the interaction strength
increases. The value of $\omega_s$ varies slowly with changes 
in the interaction strength $C$, see figure \ref{figure3}. It is 
worth noting that the simple  
formula (\ref{acoustic}) gives a good estimate of the frequency spectrum. 
For example, for $C=1000$  $\omega_s \approx 0.57$  in good agreement 
with the results shown (figure \ref{figure3}) while for $C =4300$ the value is 
$\omega_s \approx 0.78$. The relation (\ref{sound}) is quite general and 
can by applied to any toroidal geometry. For a large radius but 
tightly confined torus, so that $\langle r^{-2} \rangle \approx r_K^{-2}$ 
the relation simplifies to the 1D result as expected:
\begin{equation}
\omega = |m| \sqrt{2C/\Omega}\  r_K^{-1}
\end{equation}
where the wavenumber is quantized according to $|m|/r_K$, and the speed 
of sound is proportional to the square root of the atom density. 
One might assume, for a given radial and axial mode,  
that this pattern of frequencies for the $m$ sublevels would be repeated 
for the higher frequency bands. 
In fact this is 
not the case as confirmed  in figures \ref{figure2} and \ref{figure3}. The bands become mixed and 
the  $\omega \sim  |m|$ relation does not hold. 
A more extensive list of the frequencies 
including the highly excited modes is given 
in Table \ref{table2}.

\begin{figure}[htbp]
\begin{center}
\includegraphics[width=8cm,angle=0]{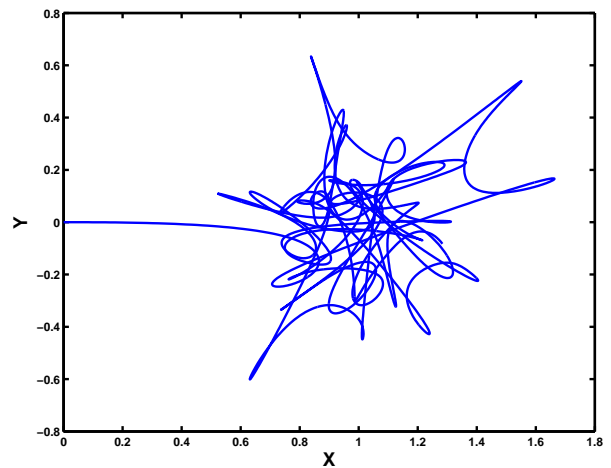}  
\caption{\label{figure5} 
Center-of-mass trajectory of a toroidal condensate current $\kappa=1$ displaced
from the trap center. 
The torus has parameters $\lambda = \sqrt{7}$, $\sigma = 2 \sqrt{\hbar/2 m_a\omega_r}$ 
and $h_0 = 50 \hbar \omega_r$ and interaction strength $C= 1000$. The 
figure shows the path traced out over a total 
time $T=100$ in the horizontal plane by the  condensate 
center-of-mass following a small horizontal displacement of the trap 
axis $\varepsilon = 1$. The combination of the $x$-dipole and $y$-dipole oscillations leading to 
precession of the condensate. These dipoles modes are degenerate and nearly in phase. 
The spectrum density of the $x$-dipole oscillations is shown in 
figure \ref{figure7}
}
\end{center}
\end{figure}

\subsection{Toroidal current flow}

\begin{figure}[htbp]
\begin{center}
\includegraphics[width=8cm,angle=0]{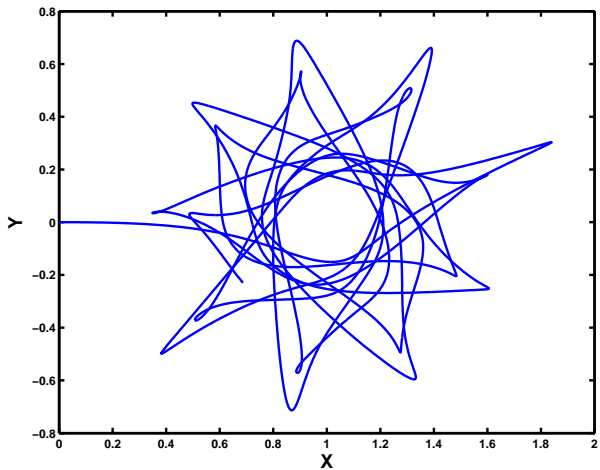}  
\caption{\label{figure6} 
Center-of-mass trajectory of a toroidal condensate (current $\kappa=1$) after 
radial displacement. 
As in figure \ref{figure5} the torus has parameters $\lambda = \sqrt{7}$, $\sigma = 2 \sqrt{\hbar/2 m_a\omega_r}$ 
and $h_0 = 50 \hbar \omega_r$, however the 
atom number is much higher in this case:$C= 4300$. The 
figure shows the path traced out over a time $T=100$ in the horizontal plane by the  condensate 
center-of-mass following a small horizontal displacement of the trap 
$\varepsilon = 1$. The center of mass performs a circular motion due to two effects:
the $x$ and $y$ dipole states are not degenerate in frequency and the amplitudes are 
out of phase, by roughly $\sim \pi/2$. 
}
\end{center}
\end{figure}

If the condensate itself has a current $\kappa \neq 0$, then the central barrier acts as 
a pinning site for the vortex. Furthermore  the $\pm |m|$  degeneracy
of the excitation modes is removed. Perturbations of the external potential are of interest 
since it allows us to study imperfections in the toroidal trap potential, and also 
the dissipation of the ring current. We propose that these modes 
can be studied by a lateral displacement of the trap. In Figure  \ref{figure4}
the spectrum of excitations of a singly-quantized  current loop $\kappa=1$ are 
presented. In this figure we consider the transition from a conventional  pancake 
geometry $h_0=0$, through a semi-toroidal trap $h_0=10$, before reaching the 
torus shape $h_0=50$. The presence of the barrier 
creates a narrower radial well, so that the $m=0$ radial excitations are pushed to higher frequencies.
The axial $m=0$ dipole mode $\omega= \lambda \approx 2.65$, is unaffected of course.
The splitting of the $m$-degeneracy 
 is greatest for the spheroidal $h_0=0$  case, in particular the 
 lowest pair of $|m|=1$ modes. The co-rotating mode $m=+1$ has a positive 
 frequency lying close to the lowest $m=2$ mode. The 
anomalous or counter-rotating mode  $m=-1$ has a negative frequency, but positive norm. 
Anomalous modes arise 
when a condensate has a stable topological excitation such 
as a vortex, indeed the presence 
of a vortex is a necessary condition for an anomalous excitation 
as its presence is required 
to satisfy conservation of energy and momentum in
the system \cite{fede01}. In this case the central barrier raises the anomalous mode to 
positive frequencies effectively creating a stable pinning site for toroidal flow. 
The stability of high 
circulation current in toroidal traps has been discussed in detail by 
Busch et al. \cite{busc01}. 
The  splitting of the $|m|=1$  pair reduces as the barrier rises and the 
condensate becomes toroidal. The closing of the energy gap reflects the fact that the 
barrier expels condensate density from the center so that the $m=-1 $ cannot 
occupy the vortex core region and create a large energy gap.  
The $|m|=1$  modes will be evident when dipole 
excitations of the circulating flow are discussed.  The $m=\pm 2$ states, 
degenerate in figure \ref{figure3} , are also split by the condensate flow and 
in the toroidal limit, $h_0=50$. The energy splittings are proportional to $|m|$ the current 
momentum. This follows from the Bogoliubov equations when the centrifugal energy terms can be 
considered as perturbations. First-order perturbation theory applied to equations 
(\ref{bogo1}) and (\ref{bogo2})  gives the energy shift:
\begin{equation}
\Delta \hbar\omega  \approx {\hbar^2 (m+\kappa)^2 \over 2m_a} (u_0,r^{-2}u_0) 
+  {\hbar^2 (m-\kappa)^2 \over 2m_a} (v_0,r^{-2}v_0) 
\end{equation} 
where $u_0$ and $v_0$ are the unperturbed quasiparticle states. This
gives the energy splitting $\Delta E = \hbar (\omega_+-\omega_-)$:
\begin{equation}
\Delta E  \approx 4 |m| \kappa  \times { \hbar^2  \over 2 m_a} \int (u_0^*u_0-v_0^*v_0)r^{-2}\
d^3{\bm r} 
\end{equation} 
 If the condensate and excitations
are tightly confined around the radius $r_K$ then:
\begin{equation}
\Delta E  \approx 4 |m| \kappa  \times { \hbar^2  \over 2 m_a r_K^2} 
\label{split}
\end{equation} 
The splitting decreases as $C$, and hence $r_K$  increases. Referring to Table \ref{table1} 
for $h_0=50$, the $|m|=1$ splitting is close to that given by the perturbation expression 
(equation \ref{split}) $\Delta E \approx 0.19$. However the lowest $|m|=2$ modes in Table \ref{table1} 
have a splitting (0.37) very slighty smaller than the pertubation result  $0.38$.

\begin{figure}[!tbp]
\centering
\includegraphics[width=7cm]{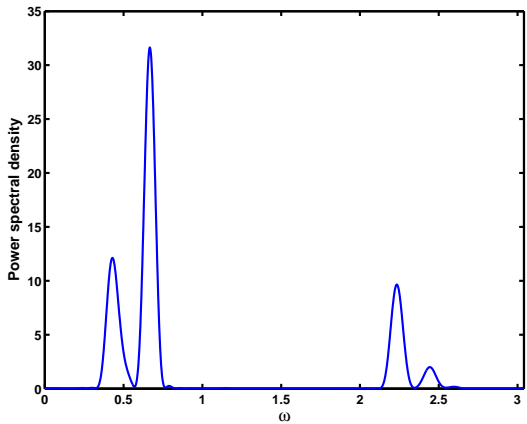} 
\caption{
\label{fig:ncU10spec} 
Spectral density of $x$-dipole of a condensate with circulation $\kappa=1$ 
corresponding to figure \ref{figure5}:
$P_{x}\left(\omega\right)$
The torus parameters are 
$\lambda = \sqrt{7}$, 
$\sigma = 2 \sqrt{\hbar/2 m_a\omega_r}$ and $h_0 = 50 \hbar \omega_r$.
The condensate interaction strength is $C= 1000$ with trap displacement
$\epsilon = 1$. The figure shows the dominance of three modes, the low
 frequency $m=\pm 1$ doublet, 
and an excited  $m=-1$ mode. The 
center of the doublet corresponds to the frequency $\omega_s= 
\sqrt{2C \langle r^{-2}\rangle} \approx 0.58$. In the graph the frequency peaks are  
shifted from the linear response results given in  Table \ref{table1} and have a 
broader splitting than the Bogoliubov approximation. The 
$x$-dipole modes, shown above, have frequencies 
$\omega_x = 0.421, 0.666, 2.237, 2.441$ while the 
corresponding $y$-dipole modes have frequencies:
$\omega_y =0.449, 0.657, 2.234, 2.422$.}
\label{figure7}
\end{figure}    

\begin{figure}[!tbp]
\begin{center}
\includegraphics[width= 7cm]{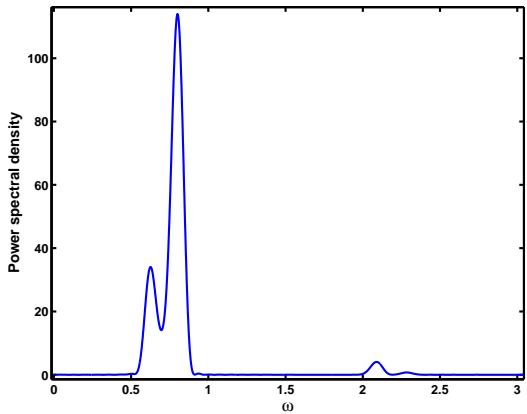} 	
\caption{\label{figure8}
Spectral density of $x$-dipole of condensate with circulation $\kappa=1$
corresponding to figure \ref{figure6}:
$P_{x}\left(\omega\right)$
The torus parameters are 
$\lambda = \sqrt{7}$, 
$\sigma = 2 \sqrt{\hbar/2 m_a\omega_r}$ and $h_0 = 50 \hbar \omega_r$.
The condensate interaction strength is $C= 4300$ with trap displacement
$\epsilon = 1$.
 The center of the doublet corresponds to the frequency $\omega_s= 
\sqrt{2C \langle \phi^2 r^{-2}\rangle} \approx 0.78$
 The figure shows the shifting upwards of the mode frequencies and 
a narrowing of the $m=\pm 1$ doublet splitting consistent with an expanded condensate with a higher 
sound speed. The modes shown above have frequencies: $ \omega_{x}  =0.625,  0.804, 2.095,  2.284$. 
The corresponding $y$ dipole modes, not shown, have similar strengths but with 
frequencies $ \omega_{y} = 0.657,  0.792, 2.086,  2.275$.}
\end{center}
\end{figure}   

 The simplest experimental 
scheme to observe these modes is to displace the potential 
in the horizontal plane and follow the motion of the condensate center-of-mass.  
Consider a sudden  adjustment of the 
trapping potential 
\begin{equation} 
V_{\rm ext }\left(x,y,z\right) \rightarrow V_{\rm ext}\left(x - \varepsilon,y,z\right)
\label{dp2} 
\end{equation}
The effect on the center-of-mass is shown in figures \ref{figure5} 
and figure \ref{figure6} in which the trajectory is traced out as 
time evolves. These graphs were obtained by direct 
solution of the time-dependent Gross-Pitaevskii
equation, following from taking 
arbitrary variation of $\psi^*$ in (\ref{vp}):
\begin{equation}
 \left[ H_0 + g |\psi|^2 -i\hbar\partial_t \right]\psi =0
\label{nlse}
\end{equation}
For very small perturbations the nonlinearity is small and 
the time-dependent linear response method (TDLR) is equivalent to the 
stationary Bogoliubov equation method outlined above. This 
employs a direct numerical solution, 
in combination with spectral analysis can also
be used to determine the frequency spectrum \cite{mcpe02} and the 
strength of each frequency component, that is the population of modes 
and density fluctuations of the collective excitations.
This can be done efficiently and accurately using spectral 
methods. The initial state can be determined 
by imposing the phase according to the value of $\kappa$ and evolving in imaginary
time  \cite{mcpe02}.
 The calculations presented here were performed on a $64 \times 64 \times 64$ 
grid. The scaling of 
grid spacing  according to the degree of confinement 
ensures maximum spatial resolution of the
condensate, so that the $z$ dimension is more tightly confined. 
The time step chosen, $\Delta t \approx 0.001 \times 2\pi/\omega_r $, is 
dictated by the characteristic timescale of the excitations. 
 A power-spectral-density 
estimate which used a Kaiser windowing function was employed to analyse the  various  
moments of interest in terms of its component frequencies. For example the 
$x$-dipole moment can be transformed as follows:
\begin{equation} 
P_{x}\left(\omega\right) = \left|\int_0^{T} e^{-i\omega t}\langle x \rangle  dt\right|^2 
\label{psd}
\end{equation}   
The spectral data obtained for a very small trap displacement $\varepsilon=0.02$  
confirms the mode  frequencies from the Bogoliubov-de Gennes equation, and establishes
the accuracy of the method. As shown in table \ref{table1}, the frequencies 
using this method are accurate 
to within a few per cent. However a realistic measurement 
process based on density imaging requires a much larger amplitude motion in order 
to resolve the fluctuations. In this case, the nonlinear effect can be important. 
In our scheme, which we propose as a viable method to measure the spectrum,  we consider 
the trap displaced by a substantial amount $\varepsilon=1$ in order to resolve the oscillations 
in the plane.
When circulation of the toroidal current is disturbed
in this manner the effect is to create large counter-circulating  currents. If 
these currents are in phase and of equal magnitude the condensate would execute a pedulum motion. 
In general the motion will be 2D if the symmetry is broken. In terms of 
experimental observables, the effect appears as oscillations of the center of mass
 with a precessional motion.
Similar precession motion for quadrupole excitation of a spheroidal vortex state has
been noted \cite{mcpe02}. The locus of the center-of-mass in the horizontal plane from 
$t=0$ to $t=100$, is 
shown in figure \ref{figure5} for $C=1000$ and $h_0/\hbar\omega_0=10$. Initially the condensate 
moves from $x=0,y=0$ towards the equilibrium point $x=1,y=0$, however this motion is converted 
an irregular  precessing  pendulum motion.  
The rate of precession is proportional to the  splitting of levels. However the center-of-mass 
motion is not simple. A clearer understanding of the 
motion follows from a spectral analysis of the $x$ and $y$ dipole moments.
In fact, the apparently irregular motion  
is dominated by only a few components the low frequency dipoles.  
In both the case $h_0=10$ (figure \ref{figure7}) 
and $h_0=50$ (figure \ref{figure8}) the $m=+1$ mode is stronger 
than the mode $m=-1$ and 
this dictates a motion with positive helicity, that is the condensate precession is in the same 
sense as the flow. However the frequencies are significantly different 
from the linear model and the $x$-$y$ degeneracy is removed due to 
the nonlinearity of the response in the $x$ direction. 
In figure \ref{figure7}, the torus parameters are $\lambda = \sqrt{7}$, 
$\sigma = 2 \sqrt{\hbar/2 m_a\omega_r}$ and $h_0 = 50 \hbar \omega_r$.
The condensate interaction strength is $C= 1000$ with trap displacement
$\epsilon = 1$. Comparing the $x$ frequencies with the values of table \ref{table1} 
we note the $y$-dipole frequencies are very similar in both cases, but the 
$x$-dipole pair splitting is increased when the trap displacement is this large. 
The motion shown in figure \ref{figure5} is a superposition of a faster $x$ dipole 
with a slower $y$ dipole. The results for the higher interaction strength $C=4300$  
show the splitting of the $|m|=1$ lines narrowing as the size of the condensate 
increases. 
The figure shows the shifting upwards of the mode frequencies and 
a narrowing of the $m=\pm 1$ doublet splitting consistent with an expanded condensate with a higher 
sound speed. The modes shown have frequencies: $ \omega_{x}  =0.625,  0.804, 2.095,  2.284$. 
The corresponding $y$ dipole modes have similar strengths but with 
frequencies $ \omega_{y} = 0.657,  0.792, 2.086,  2.275$.
The nonlinear interactions give different splittings for the $x$ and 
$y$ dipoles: $\delta \omega_x = 0.179$ compared 
with $\delta \omega_y = 0.135$ both of which are slightly smaller than the linear 
perturbation theory  (\ref{split}) $\delta \omega \approx 0.200$. The
major contrast between figures \ref{figure5} and \ref{figure6} is the 
orbiting motion observed for the higher value of $C$, although the frequency 
components shown in figures \ref{figure7} and \ref{figure8}  are of similar 
relative strengths, the $x$ and $y$ dipoles have a large phase difference 
close to $\pi/2$ for $C=4300$ and explains the more circular  character 
of the motion. This is slightly surprising in that the usual behavior of 
collective mode of oscillations is that the 
frequency dependence on $N$ is relatively 
weak for large $N$. In this  respect, collective excitation frequencies 
are usually a poor method of estimating condensate fraction compared 
with the hydrostatic pressure. However, in this case the toroidal geometry 
seems to retain the quantum features of the excitations even for 
large particle number.

\section{Conclusions}
\label{sec5}

In this paper we studied  the 
spectrum of collective excitations of oblate toroidal condensates 
within the Bogoliubov approximation, and 
explored the dynamical stability of ring currents around the torus.
The main features we noted are generic to this design of trap and would 
apply to similar geometries and would be exhibited in experiments 
on toroidal condensates.
The transition from 
spheroidal to toroidal geometry of the trap displaces the energy
 levels into narrow bands. 
The lowest-order modes are quasi one-dimensional circulations with dispersion 
relation $\omega \sim m \omega_s $ with $m$ quantized.  
When the condensate has an  toroidal current flow, the low-energy circulations 
are split into co-rotating and counter-rotating pairs. 
 A simple, but accurate,  
 formula is presented for the lowest angular acoustic 
 modes of excitation, and the splitting energy when a background
 current is present. The stability of the flow is maintained by the 
 central barrier acting as a pinning site.

Instabilities  will  most readily occur by radial dipole excitations. 
The dipole states are nearly degenerate in the toroidal 
limit.  Small condensate currents 
when disturbed create slow precessional motion through the 
center of the trap through mixing of the degenerate $x$ and $y$ dipole 
states. For a tightly confined torus, 
these modes are well separated in energy and accessible to 
observation. We propose an experiment that would detect these modes. A realistic 
measurement requires resolution of the global condensate motion. 
Large displacements, such as those we propose, give rise 
to  small nonlinear shifts in the frequencies beyond the Bogoliubov model. 
We also note that the highly occupied condensate performs a  circular orbiting motion of the
center-of-mass. Such features are resolvable and experimentally measurable
with current technology and expertise.

\bibliography{torus_paper_pra}

\begin{thebibliography}{22}
\expandafter\ifx\csname natexlab\endcsname\relax\def\natexlab#1{#1}\fi
\expandafter\ifx\csname bibnamefont\endcsname\relax
  \def\bibnamefont#1{#1}\fi
\expandafter\ifx\csname bibfnamefont\endcsname\relax
  \def\bibfnamefont#1{#1}\fi
\expandafter\ifx\csname citenamefont\endcsname\relax
  \def\citenamefont#1{#1}\fi
\expandafter\ifx\csname url\endcsname\relax
  \def\url#1{\texttt{#1}}\fi
\expandafter\ifx\csname urlprefix\endcsname\relax\def\urlprefix{URL }\fi
\providecommand{\bibinfo}[2]{#2}
\providecommand{\eprint}[2][]{\url{#2}}

\bibitem[{\citenamefont{Penckwitt and Ballagh}(2001)}]{penc01}
\bibinfo{author}{\bibfnamefont{A.~A.} \bibnamefont{Penckwitt}}
  \bibnamefont{and} \bibinfo{author}{\bibfnamefont{R.~J.}
  \bibnamefont{Ballagh}}, \bibinfo{journal}{J. Phys. B: At. Mol. Opt. Phys.}
  \textbf{\bibinfo{volume}{34}}, \bibinfo{pages}{1523} (\bibinfo{year}{2001}).

\bibitem[{\citenamefont{Hutchinson and Zaremba}(1998)}]{hutc98}
\bibinfo{author}{\bibfnamefont{D.~W.} \bibnamefont{Hutchinson}}
  \bibnamefont{and} \bibinfo{author}{\bibfnamefont{E.}~\bibnamefont{Zaremba}},
  \bibinfo{journal}{Phys. Rev. A} \textbf{\bibinfo{volume}{57}},
  \bibinfo{pages}{1280} (\bibinfo{year}{1998}).

\bibitem[{\citenamefont{Rokhsar}()}]{rokh97}
\bibinfo{author}{\bibfnamefont{D.~S.} \bibnamefont{Rokhsar}},
  \eprint{cond-mat/9709212}.

\bibitem[{\citenamefont{Salasnich et~al.}(1999)\citenamefont{Salasnich, Parola,
  and Reatto}}]{sala99}
\bibinfo{author}{\bibfnamefont{L.}~\bibnamefont{Salasnich}},
  \bibinfo{author}{\bibfnamefont{A.}~\bibnamefont{Parola}}, \bibnamefont{and}
  \bibinfo{author}{\bibfnamefont{L.}~\bibnamefont{Reatto}},
  \bibinfo{journal}{Phys. Rev. A} \textbf{\bibinfo{volume}{59}},
  \bibinfo{pages}{2990} (\bibinfo{year}{1999}).

\bibitem[{\citenamefont{Benakli et~al.}(1999)\citenamefont{Benakli, Raghavan,
  Smerzi, Fantoni, and Shenov}}]{bena99}
\bibinfo{author}{\bibfnamefont{M.}~\bibnamefont{Benakli}},
  \bibinfo{author}{\bibfnamefont{S.}~\bibnamefont{Raghavan}},
  \bibinfo{author}{\bibfnamefont{A.}~\bibnamefont{Smerzi}},
  \bibinfo{author}{\bibfnamefont{S.}~\bibnamefont{Fantoni}}, \bibnamefont{and}
  \bibinfo{author}{\bibfnamefont{S.~R.} \bibnamefont{Shenov}},
  \bibinfo{journal}{Europhys. Lett.} \textbf{\bibinfo{volume}{46}},
  \bibinfo{pages}{275} (\bibinfo{year}{1999}).

\bibitem[{\citenamefont{Arnold and Riis}(2002)}]{arno02}
\bibinfo{author}{\bibfnamefont{A.~S.} \bibnamefont{Arnold}} \bibnamefont{and}
  \bibinfo{author}{\bibfnamefont{E.}~\bibnamefont{Riis}}, \bibinfo{journal}{J.
  Mod. Opt.} \textbf{\bibinfo{volume}{49}}, \bibinfo{pages}{959}
  (\bibinfo{year}{2002}).

\bibitem[{\citenamefont{Crompvoets et~al.}(2001)\citenamefont{Crompvoets,
  Bethlem, Jongma, and Meijer}}]{crom01}
\bibinfo{author}{\bibfnamefont{F.~M.} \bibnamefont{Crompvoets}},
  \bibinfo{author}{\bibfnamefont{H.~L.} \bibnamefont{Bethlem}},
  \bibinfo{author}{\bibfnamefont{R.~T.} \bibnamefont{Jongma}},
  \bibnamefont{and} \bibinfo{author}{\bibfnamefont{G.}~\bibnamefont{Meijer}},
  \bibinfo{journal}{Nature} \textbf{\bibinfo{volume}{411}},
  \bibinfo{pages}{174} (\bibinfo{year}{2001}).

\bibitem[{\citenamefont{Drummond et~al.}(2001)\citenamefont{Drummond,
  Eleftheriou, Huang, and Kheruntsyan}}]{drum01}
\bibinfo{author}{\bibfnamefont{P.~D.} \bibnamefont{Drummond}},
  \bibinfo{author}{\bibfnamefont{A.}~\bibnamefont{Eleftheriou}},
  \bibinfo{author}{\bibfnamefont{K.}~\bibnamefont{Huang}}, \bibnamefont{and}
  \bibinfo{author}{\bibfnamefont{K.~V.} \bibnamefont{Kheruntsyan}},
  \bibinfo{journal}{Phys. Rev. A} \textbf{\bibinfo{volume}{63}},
  \bibinfo{pages}{053602} (\bibinfo{year}{2001}).

\bibitem[{\citenamefont{Garay et~al.}(2001)\citenamefont{Garay, Anglin, Cirac,
  and Zoller}}]{gara01}
\bibinfo{author}{\bibfnamefont{L.~J.} \bibnamefont{Garay}},
  \bibinfo{author}{\bibfnamefont{J.~R.} \bibnamefont{Anglin}},
  \bibinfo{author}{\bibfnamefont{J.~I.} \bibnamefont{Cirac}}, \bibnamefont{and}
  \bibinfo{author}{\bibfnamefont{P.}~\bibnamefont{Zoller}},
  \bibinfo{journal}{Phys. Rev. A} \textbf{\bibinfo{volume}{63}},
  \bibinfo{pages}{023611} (\bibinfo{year}{2001}).

\bibitem[{\citenamefont{Brand and Reinhardt}(2001)}]{bran01}
\bibinfo{author}{\bibfnamefont{J.}~\bibnamefont{Brand}} \bibnamefont{and}
  \bibinfo{author}{\bibfnamefont{W.~P.} \bibnamefont{Reinhardt}},
  \bibinfo{journal}{J. Phys. B: At. Mol. Opt. Phys.}
  \textbf{\bibinfo{volume}{34}}, \bibinfo{pages}{L113} (\bibinfo{year}{2001}).

\bibitem[{\citenamefont{Petrov et~al.}(2000)\citenamefont{Petrov, Shlyapnikov,
  and Walraven}}]{petr00}
\bibinfo{author}{\bibfnamefont{D.~S.} \bibnamefont{Petrov}},
  \bibinfo{author}{\bibfnamefont{G.~V.} \bibnamefont{Shlyapnikov}},
  \bibnamefont{and} \bibinfo{author}{\bibfnamefont{J.~T.~M.}
  \bibnamefont{Walraven}}, \bibinfo{journal}{Phys. Rev. Lett.}
  \textbf{\bibinfo{volume}{85}}, \bibinfo{pages}{3745} (\bibinfo{year}{2000}).

\bibitem[{\citenamefont{Girardeau and Wright}(2001)}]{gira01}
\bibinfo{author}{\bibfnamefont{M.~D.} \bibnamefont{Girardeau}}
  \bibnamefont{and} \bibinfo{author}{\bibfnamefont{E.~M.}
  \bibnamefont{Wright}}, \bibinfo{journal}{Phys. Rev. Lett.}
  \textbf{\bibinfo{volume}{87}}, \bibinfo{pages}{210401}
  (\bibinfo{year}{2001}).

\bibitem[{\citenamefont{Davis et~al.}(1995)\citenamefont{Davis, Mewes, Andrews,
  van Drutten, Durfree, Kurn, and Ketterle}}]{davi95}
\bibinfo{author}{\bibfnamefont{K.~B.} \bibnamefont{Davis}},
  \bibinfo{author}{\bibfnamefont{M.~O.} \bibnamefont{Mewes}},
  \bibinfo{author}{\bibfnamefont{M.~R.} \bibnamefont{Andrews}},
  \bibinfo{author}{\bibfnamefont{N.~J.} \bibnamefont{van Drutten}},
  \bibinfo{author}{\bibfnamefont{D.~S.} \bibnamefont{Durfree}},
  \bibinfo{author}{\bibfnamefont{D.~M.} \bibnamefont{Kurn}}, \bibnamefont{and}
  \bibinfo{author}{\bibfnamefont{W.}~\bibnamefont{Ketterle}},
  \bibinfo{journal}{Phys. Rev. Lett.} \textbf{\bibinfo{volume}{75}},
  \bibinfo{pages}{3969} (\bibinfo{year}{1995}).

\bibitem[{\citenamefont{Matthews et~al.}(1999)\citenamefont{Matthews, Anderson,
  Haljan, Hall, Wieman, and Cornell}}]{matt99}
\bibinfo{author}{\bibfnamefont{M.~R.} \bibnamefont{Matthews}},
  \bibinfo{author}{\bibfnamefont{B.~P.} \bibnamefont{Anderson}},
  \bibinfo{author}{\bibfnamefont{P.}~\bibnamefont{Haljan}},
  \bibinfo{author}{\bibfnamefont{D.~S.} \bibnamefont{Hall}},
  \bibinfo{author}{\bibfnamefont{C.~E.} \bibnamefont{Wieman}},
  \bibnamefont{and} \bibinfo{author}{\bibfnamefont{E.~A.}
  \bibnamefont{Cornell}}, \bibinfo{journal}{Phys. Rev. Lett.}
  \textbf{\bibinfo{volume}{83}}, \bibinfo{pages}{2498} (\bibinfo{year}{1999}).

\bibitem[{\citenamefont{Busch et~al.}()\citenamefont{Busch, Anglin, and
  Zurek}}]{busc01}
\bibinfo{author}{\bibfnamefont{T.}~\bibnamefont{Busch}},
  \bibinfo{author}{\bibfnamefont{J.}~\bibnamefont{Anglin}}, \bibnamefont{and}
  \bibinfo{author}{\bibfnamefont{W.}~\bibnamefont{Zurek}},
  \eprint{cond-mat/0103394}.

\bibitem[{\citenamefont{Schulte et~al.}(2002)\citenamefont{Schulte, Santos,
  Sanpera, and Lewenstein}}]{schu02}
\bibinfo{author}{\bibfnamefont{T.}~\bibnamefont{Schulte}},
  \bibinfo{author}{\bibfnamefont{L.}~\bibnamefont{Santos}},
  \bibinfo{author}{\bibfnamefont{A.}~\bibnamefont{Sanpera}}, \bibnamefont{and}
  \bibinfo{author}{\bibfnamefont{M.}~\bibnamefont{Lewenstein}},
  \bibinfo{journal}{Phys. Rev. A} \textbf{\bibinfo{volume}{66}},
  \bibinfo{pages}{033602} (\bibinfo{year}{2002}).

\bibitem[{\citenamefont{McPeake et~al.}(2002)\citenamefont{McPeake, Nilsen, and
  McCann}}]{mcpe02}
\bibinfo{author}{\bibfnamefont{D.}~\bibnamefont{McPeake}},
  \bibinfo{author}{\bibfnamefont{H.}~\bibnamefont{Nilsen}}, \bibnamefont{and}
  \bibinfo{author}{\bibfnamefont{J.~F.} \bibnamefont{McCann}},
  \bibinfo{journal}{Phys. Rev. A} \textbf{\bibinfo{volume}{65}},
  \bibinfo{pages}{063601} (\bibinfo{year}{2002}).

\bibitem[{\citenamefont{Nilsen et~al.}(2003)\citenamefont{Nilsen, McPeake, and
  McCann}}]{nils03}
\bibinfo{author}{\bibfnamefont{H.}~\bibnamefont{Nilsen}},
  \bibinfo{author}{\bibfnamefont{D.}~\bibnamefont{McPeake}}, \bibnamefont{and}
  \bibinfo{author}{\bibfnamefont{J.~F.} \bibnamefont{McCann}},
  \bibinfo{journal}{J. Phys. B: At. Mol. Opt. Phys.}
  \textbf{\bibinfo{volume}{36}}, \bibinfo{pages}{1703} (\bibinfo{year}{2003}).

\bibitem[{\citenamefont{Adams and Riis}(1997)}]{adam97}
\bibinfo{author}{\bibfnamefont{C.~S.} \bibnamefont{Adams}} \bibnamefont{and}
  \bibinfo{author}{\bibfnamefont{E.}~\bibnamefont{Riis}},
  \bibinfo{journal}{Prog. Quantum Elect.} \textbf{\bibinfo{volume}{21}},
  \bibinfo{pages}{1} (\bibinfo{year}{1997}).

\bibitem[{\citenamefont{Dalfovo et~al.}(1999)\citenamefont{Dalfovo, Giorgini,
  Pitaevskii, and Stringari}}]{dalf99}
\bibinfo{author}{\bibfnamefont{F.}~\bibnamefont{Dalfovo}},
  \bibinfo{author}{\bibfnamefont{S.}~\bibnamefont{Giorgini}},
  \bibinfo{author}{\bibfnamefont{L.~P.} \bibnamefont{Pitaevskii}},
  \bibnamefont{and}
  \bibinfo{author}{\bibfnamefont{S.}~\bibnamefont{Stringari}},
  \bibinfo{journal}{Rev. Mod. Phys.} \textbf{\bibinfo{volume}{71}},
  \bibinfo{pages}{463} (\bibinfo{year}{1999}).

\bibitem[{\citenamefont{Golub and van Loan}(1996)}]{golu96}
\bibinfo{author}{\bibfnamefont{G.~H.} \bibnamefont{Golub}} \bibnamefont{and}
  \bibinfo{author}{\bibfnamefont{C.~F.} \bibnamefont{van Loan}},
  \emph{\bibinfo{title}{Matrix Computations}} (\bibinfo{publisher}{Johns
  Hopkins University Press}, \bibinfo{year}{1996}).

\bibitem[{\citenamefont{Feder et~al.}(2001)\citenamefont{Feder, Svidzinsky,
  Fetter, and Clark}}]{fede01}
\bibinfo{author}{\bibfnamefont{D.~L.} \bibnamefont{Feder}},
  \bibinfo{author}{\bibfnamefont{A.~A.} \bibnamefont{Svidzinsky}},
  \bibinfo{author}{\bibfnamefont{A.~L.} \bibnamefont{Fetter}},
  \bibnamefont{and} \bibinfo{author}{\bibfnamefont{C.~W.} \bibnamefont{Clark}},
  \bibinfo{journal}{Phys. Rev. Lett.} \textbf{\bibinfo{volume}{86}},
  \bibinfo{pages}{564} (\bibinfo{year}{2001}).

\end{thebibliography}

\end{document}